\documentclass[pre,twocolumn,showpacs,floatfix]{revtex4}
\usepackage{graphicx}
\usepackage{amsmath}
\usepackage{amssymb}
\usepackage{subfigure}
\usepackage{multirow}
\usepackage{psfrag}
\newcommand{\w}[1]{\vec{{#1}}}
\newcommand{\ww}[1]{\underline{\underline{{\bf #1}}}}
\newcommand{\sig}{\ww{\sigma}}
\newcommand{\Sig}{\ww{\Sigma}}

\newcommand{\be}{\begin{equation}}
\newcommand{\ee}{\end{equation}}
\newcommand{\bea}{\begin{eqnarray}}
\newcommand{\eea}{\end{eqnarray}}
\newcommand{\ba}{\begin{aligned}}
\newcommand{\ea}{\end{aligned}}
\newcommand{\bma}{\begin{bmatrix}}
\newcommand{\ema}{\end{bmatrix}}

\newcommand{\bi}{\begin{itemize}}
\newcommand{\ei}{\end{itemize}}

\newcommand{\tr}{\text{Tr}\hspace{.05em}}

\newcommand{\ave}[1]{\langle #1 \rangle}

\newcommand{\qs}{quasistatic}
\newcommand{\testtrois}{shear}
\newcommand{\Testtrois}{Shear}
\newcommand{\sigI}{\sigma_{\rm I}}
\newcommand{\sigII}{\sigma_{\rm II}}
\newcommand{\sigIII}{\sigma_{\rm III}}
\renewcommand{\imath}{i}
\newcommand{\sfab}{S_{\rm fab}}
\newcommand{\sfor}{S_{\rm for}}

\begin{document}

\title{Solidlike behavior and anisotropy in rigid frictionless bead assemblies}
\author{Pierre-Emmanuel Peyneau}
\email{pierre-emmanuel.peyneau@lcpc.fr}
\author{Jean-No\"el Roux}
\affiliation{Universit\'e Paris-Est, UR Navier,
LMSGC\footnote{ Laboratoire des Mat\'eriaux et des Structures du G\'enie Civil is a joint laboratory depending on
 Laboratoire Central des Ponts et Chauss\'ees, \'Ecole Nationale des Ponts et Chauss\'ees and Centre National de la Recherche Scientifique},
2 all\'ee Kepler, Cit\'e Descartes, 77420 Champs-sur-Marne, France}
\date{Accepted for publication in Phys. Rev. E, October 1, 2008}

\begin{abstract}
We investigate the structure and mechanical behavior of assemblies of frictionless, nearly rigid equal-sized beads, in the quasistatic limit, 
by numerical simulation.  Three different loading paths are explored: triaxial compression, triaxial extension and simple shear. 
Generalizing recent result, we show that the material, despite rather strong finite sample size effects, 
is able to sustain a finite deviator stress in the macroscopic limit, along all three paths, without dilatancy. 
The shape of the yield surface in principal stress space differs somewhat from the Mohr-Coulomb prediction, and is more adequately described by the 
Lade-Duncan or Matsuoka-Nakai criteria. We study geometric characteristics and force networks under varying stress levels within the supported range. 
Although the scalar state variables stay equal to the values observed in systems under isotropic pressure, the material, once subjected to a deviator stress, 
possesses some fabric and force distribution anisotropies. Each kind of anisotropy can be described, in good approximation, by a single parameter. 
Within the supported stress range, along each one of the three investigated stress paths, among those three quantities: deviator stress to mean stress ratio, 
fabric anisotropy parameter, force anisotropy parameter, any one determines the values of the two others. The pair correlation function also exhibits 
short range anisotropy, up to a distance between bead surfaces of the order of 10\% of the diameter. The tensor of elastic moduli is shown to possess a nearly singular, 
uniaxial structure related to stress anisotropy. Possible stress-strain relations in monotonic loading paths are also discussed.
\end{abstract}


\maketitle

\section{Introduction\label{sec:intro}}
The disordered packing of rigid, frictionless spherical balls epitomizes the large class of materials made of 
athermal, amorphous assemblies of particles with extremely short range interactions,
such as granular materials~\cite{JNB96,HHL98,degennes99,HW04,GRMH05}, concentrated suspensions~\cite{SP05,OBR06}, or some glasses~\cite{Debenedetti}. 
Obviously a highly idealized material, it is perhaps exclusively studied by numerical simulations~\cite{MGJS99,SEGHL02,OSLN03,DTS05,iviso1,PR08a}.
However, its main merit is to capture the essential role of steric exclusion and
packing geometry in the rheological properties of many different materials (termed ``jammed''~\cite{LN01} in the recent literature). 
In general, ``jammed'' particulate systems with strongly repulsive interactions tend to behave like plastic solids for nearly isotropic stress states, 
and to flow like liquids once the deviatoric stress reaches some failure threshold. 
The solidlike regime of granular materials has long been described and modeled 
at the continuum scale in the field of soil mechanics~\cite{HHL98,DMWood,MIT93}. 

Assemblies of frictionless and cohesionless grains, in the rigid limit, have two remarkable properties~\cite{JNR2000}.
First, equilibrium configurations under specified externally applied loads minimize potential energy, thereby satisfying geometric optimization
criteria. In particular, equilibrium states under isotropic pressure realize a local maximum of density in configuration space, subject to impenetrability
constraints. One thus obtains, with assembling procedures fast enough to bypass incipient crystallisation, the so-called random close packing (RCP) states
of sphere packings~\cite{OSLN03,DTS05,iviso1}, with solid fraction $\Phi\simeq 0.64$. 
Then, the force-carrying contact network (the backbone) is generically devoid of force indeterminacy, and even isostatic with circular or 
spherical objects~\cite{JNR2000,OSLN03,DTS05}. Consequently, equilibrium forces are geometrically determined, as well as the load increments necessary to 
destabilize contact networks; and such materials, in the solid state, tend to deform in a sequence of rearranging events, in which the contact structure
gets continuously broken and repaired~\cite{JNR2000,TW00,CR2000}. 

In spite of those appealing properties of frictionless spheres (or disks in 2D), 
which highlight the connections between geometry and mechanics and endow them with quite generic features, 
the study  of those model materials is still incomplete in the published literature. Numerical investigations have mostly focused on 
the geometry of RCP states~\cite{OSLN03,DTS05,iviso1}, 
on the possible effects of confining pressure variations~\cite{MJS00,iviso2} and specific elastic properties~\cite{SiLiNa06,iviso3}, on the one hand; 
and on steady-state shear flows~\cite{Dacruz05,XO06,Hatano07} on the other hand. The solid range, in which moderate deviator stresses are supported by 
anisotropic packings in equilibrium, has hardly been investigated.
 
In a recent publication~\cite{PR08a}, we checked that rigid, frictionless bead packings have a finite macroscopic friction coefficient $\mu^*$ 
in simple shear, and showed them to be devoid of dilatancy, unlike dense frictional grain assemblies. The non-vanishing value of $\mu^*$ was attributed to 
the possibility to form equilibrium structures with anisotropic contact networks. Both static (yield threshold) and dynamic 
(i.e., measured in steady shear flows) values of $\mu^*$ were shown to agree in the limit of large samples.

The present paper further investigates the mechanical properties of solidlike assemblies of frictionless beads under quasistatic loading conditions. 
The model material, initially assembled under an isotropic pressure, is subjected to different deviatoric loading paths (Section~\ref{sec:model}), so that
a failure criterion, or yield surface, delineating the stable solid range in stress space can be
identified in the macroscopic limit (Section~\ref{sec:failure}) --thus generalizing the friction angle measured in simple shear.
Then we study the geometric and micromechanical features of equilibrium states throughout the supported range of stresses, generalizing the results obtained
on isotropic packings to systems with various levels and various directions of anisotropy (Section~\ref{sec:before}).
Stresses are related to fabric and force distribution anisotropy
parameters by a simple formula. We show how the anomalous elastic moduli tensor of frictionless, nearly rigid networks is affected by stress anisotropy. 
A macroscopic stress-strain relation appears to be approached along the investigated monotonic loading paths, in spite of large statistical uncertainties.
The paper ends with a discussion (Section~\ref{sec:discussion}).
\section{Model material and simulated mechanical tests\label{sec:model}}

\subsection{Constituents and microscopic interactions\label{sec:micromech}}
We consider granular assemblies made of nearly rigid equal-sized beads of diameter $a$ and mass $m$, enclosed in a cuboidal simulation box.
Beads interact through their contacts: the force transmitted is purely normal and is the sum of a Hertzian elastic term:
\be
F_N^e =  \tilde{E}\sqrt{a}h^{3/2} / 3 \equiv \frac{2}{3} K_N(h) h,
\label{eq:elas}
\ee
and of a viscous term:
\be
F_N^v = \zeta (m\tilde{E})^{1/2} (ah)^{1/4} \dot{h} = \zeta \sqrt{2mK_N(h)} \dot{h}.
\label{eq:visc}
\ee
$h$ is the normal elastic deflection, $\tilde{E}$ is a notation for $E / (1-\nu^2)$, where $E$
is the Young modulus of the material the beads are made of, and $\nu$ its Poisson ratio, and $\zeta$ is the level of viscous damping. 
$K_N(h)$ is the equivalent spring constant associated with the elastic force given by Eq.~\eqref{eq:elas}. 
Noteworthily, albeit nonlinear, Eqs~\eqref{eq:elas} and \eqref{eq:visc} entail a velocity-independent normal 
restitution coefficient $e_N(\zeta)$ in binary collisions. 
All the simulations reported here have been performed with $\zeta = 0.98$ ($e_N = 3.3 \times 10^{-3}$). 
This model has already been employed and discussed in several recent publications~\cite{iviso1,PR08a}. 
Finally, as normal contact forces have no moment on spherical particles, their rotation is ignored.

\subsection{Boundary conditions and numerical tests\label{sec:numtests}}
Three different mechanical tests are numerically implemented to probe the solid behavior and the yield stress condition of the material. 
Those tests involve an external control on some of the entries of the Cauchy stress tensor $\ww{\sigma}$.
For a granular system at mechanical equilibrium, its expression involves the volume $V$ of the system, the intergranular force $\vec{F}_{ij}$
and the center-to-center vector $\vec{r}_{ij}$ for all pairs $(i,j)$ of contacting grains~\cite{BR90,CMNN81}:
\be
\ww{\sigma} = \frac{1}{V} \sum_{i<j} \vec{F}_{ij} \otimes \vec{r}_{ij} \label{eq:stresstensor}
\ee
Compressive stresses and shrinking strains are positive in our convention.

In order to avoid any side wall effect, the simulation cell has periodic boundary conditions
in all three directions (possibly affected by the Lees-Edwards procedure~\cite{AT87} when a non-diagonal stress component is imposed). 
Simulation cell edges have lengths denoted as $(L_\alpha)_{1 \leq \alpha \leq 3}$ along the three coordinate directions of orthonormal 
basis $(\vec e_\alpha))_{1 \leq \alpha \leq 3}$. 
Details on the equations governing the $L_{\alpha}$'s and the possible shear strain variable may be found in Ref.~\cite{PR08a}.

Before performing a mechanical test, an initial configuration is prepared under isotropic pressure $P$ with the same procedure 
as in~\cite{PR08a,iviso1}. A granular gas of hard spheres, 
initially positioned on an FCC lattice, is thermalized with collisions that preserve kinetic energy and then isotropically compressed
[with the dissipative mechanical model, 
Eqs~\eqref{eq:elas}-\eqref{eq:visc}]
until a mechanical equilibrium state is reached under presssure $P$. In the limit of small $P$,
these isotropic equilibrated configurations are the RCP states, as studied in Refs.~\cite{OSLN03,DTS05,iviso1}.

Once prepared, the material may be subjected to various loading paths. 
Three distinct \qs{} mechanical tests have been implemented, on externally applying stress tensor $\ww{\Sigma}$: 
(i) Axisymmetric triaxial compression (TC) test: $\Sig = \Sigma_1 \w{e}_1 \otimes \w{e}_1 + \Sigma_2 \w{e}_2 \otimes \w{e}_2 + \Sigma_3 \w{e}_3 \otimes \w{e}_3$, 
with $\Sigma_1 = \Sigma_2 < \Sigma_3$; (ii) Axisymmetric triaxial extension (TE) test : 
$\Sig = \Sigma_1 \w{e}_1 \otimes \w{e}_1 + \Sigma_2 \w{e}_2 \otimes \w{e}_2 + \Sigma_3 \w{e}_3 \otimes \w{e}_3$ with $\Sigma_1 = \Sigma_2 > \Sigma_3$; 
(iii) \Testtrois{} (S) test: $\Sig = P (\w{e}_1 \otimes \w{e}_1 + \w{e}_2 \otimes \w{e}_2 + \w{e}_3 \otimes \w{e}_3) + 
\tau (\w{e}_1 \otimes \w{e}_2 + \w{e}_2 \otimes \w{e}_1)$. 

Each test is employed to assess the material behavior in a particular direction of the principal stress space, 
which is the three-dimensional Euclidean space spanned by the stress tensor eigenvalues $\sigma_1$, $\sigma_2$, and $\sigma_3$ (the principal stresses). 
The principal stresses, if listed in decreasing order, are also denoted as $\sigI \geq \sigII \geq \sigIII$ in the sequel.
In equilibrium under the prescribed stress loading paths
(TC, TE and S tests) their values are listed in Tab.~\ref{tab:diagstr}. 

\begin{figure}
\centering
\includegraphics[width=6cm]{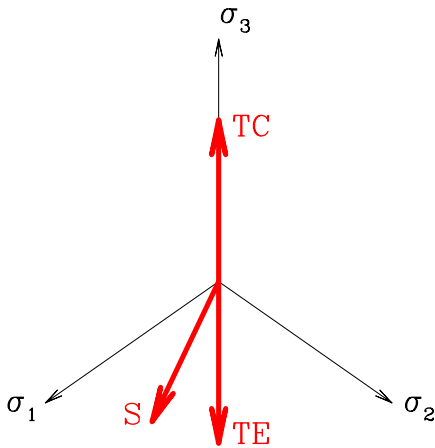}
\caption{(Color online) Sketch of the directions tested in a deviatoric plane with a triaxial compression test (with $\sigma_3 = \sigI$), 
a triaxial extension test (with $\sigma_3 = \sigIII$), and a shear test (with $(\sigma_1,\sigma_2,\sigma_3) = (\sigI,\sigIII,\sigII)$).\label{fig:loadpaths}}
\end{figure}

In all implemented tests, pressure $P= \tr{\Sig}/3$ is kept constant while deviator stress $\Sig - P\ww{1}$
is stepwise increased, with increments $\delta\Sig$. We chose to
apply $\delta\Sigma _3= 0.005\times P$ (or $-0.005\times  P$) in TC (respectively: TE) tests, whence $\delta\Sigma _1 = \delta\Sigma _2= \pm 0.0025 \times P$, and 
$\delta\tau =   0.005\times  P$ in S tests.
In principal stress space, the load therefore remains in a given deviatoric plane, i.e.,  a plane orthogonal to the trisectrix $\sigma_1 = \sigma_2 = \sigma_3$. 
The three studied stress paths are represented in Fig.~\ref{fig:loadpaths}.
\begin{table}\begin{ruledtabular}
\begin{tabular}{cccc}
Test & $\sigI$ & $\sigII$ & $\sigIII$ \\ \hline
TC   & $\sigma_{33}$ & $\sigma_{11}$ & $\sigma_{11}$ \\
TE   & $\sigma_{11}$ & $\sigma_{11}$ & $\sigma_{33}$ \\
S    & $\sigma_{33} + |\sigma_{12}|$ & $\sigma_{33}$ & $\sigma_{33} - |\sigma_{12}|$
\end{tabular}
\caption{Principal stresses for the triaxial compression, triaxial extension and \testtrois{} tests.\label{tab:diagstr}}
\end{ruledtabular}
\end{table}

\begin{figure}
\centering
\includegraphics[height=6cm]{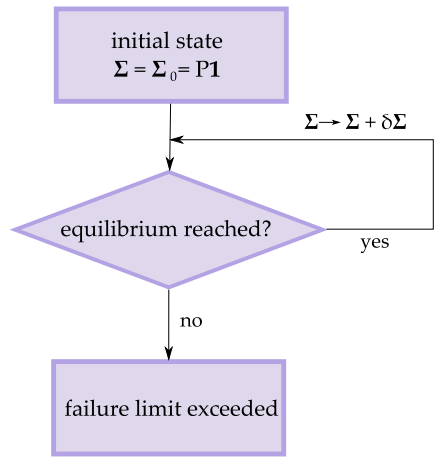}
\caption{\label{fig:step} Stepwise procedure employed to assess the failure properties of the material.}
\end{figure}

For each prescribed stress tensor $\Sig$, one waits until a satisfactory mechanical equilibrium is reached before incrementing $\Sig$.
A system is deemed equilibrated if the resultant force is zero on each bead, with a tolerance set to $10^{-4} a^2 P$, 
and $\sigma_{\alpha\beta} = \Sigma_{\alpha\beta}$ for each imposed stress component, with a relative error smaller than $10^{-4}$. 
The calculation is stopped if the packing remains out of mechanical equilibrium under the imposed stress tensor after $5\times 10^7$ time steps and a total strain of 10\%. 
The last value of $\Sig$ for which an equilibrium state was reached is kept as an estimate of the failure threshold. 
This procedure is schematized in Fig.~\ref{fig:step}.
The same simulations are carried out on a number of different randomly assembled initial configurations to achieve statistical representativity. Numerical results
are averages over available samples, the error bars shown on the figures extending to one r.m.s. sample-to-sample deviation on each side of the mean value.

\subsection{Dimensionless control parameters\label{sec:param}}
Simulation results depend on a small set of dimensionless numbers, which combine material properties and mechanical test parameters.
Most definitions recalled below are the same as in Refs.~\cite{PR08a,iviso1}. 

The stiffness parameter, defined as
\[
\kappa = \left(\frac{\tilde{E}}{P}\right)^{2/3},
\]
is such that $\kappa^{-1}$ measures the typical elastic deflection relative to particle diameter, $h/a$. Most simulations are conducted with
$\kappa = 3.9\times 10^4$ (corresponding to glass beads under $P=10$~kPa~\cite{iviso1,PR08a}). 
Half of the S tests have also been conducted with $\kappa = 8.4\times 10^3$.
From~\cite{PR08a}, we know that with such stiffness levels, the difference between the various observables measured in the simulations 
and their values in the rigid limit of $\kappa\to +\infty$ is smaller than the statistical uncertainty.

Although some dissipation is necessary in the model to reach mechanical equilibrium,
the level of viscous damping $\zeta$ is irrelevant in the \qs{} limit~\cite{PR08a}.
Here it is set to 0.98, whence a low restitution coefficient, $e_N = 3.3 \times 10^{-3}$.

When the cell is being deformed, with strain rate $\dot{\epsilon}$, 
the importance of inertia effects is characterized by the inertial number $I$, defined, as in Refs.~\cite{Gdr04,Dacruz05,Hatano07,PR08a}, by
\[
I = \dot{\epsilon}\sqrt{\frac{m}{Pa}}
\]
The \qs{} limit corresponds to $I \rightarrow 0$. 
In order to avoid excessive acceleration of the system, a control on  $\dot{\epsilon}$ is enforced, like in~\cite{iviso1,iviso2}, 
so that $I$ never exceeds $10^{-4}$.

Ratio $\delta\Sigma / P$ of deviator step to pressure is another control parameter, 
which should be kept to small values to track the quasistatic evolution of the system as accurately as possible.
The values given in Sec.~\ref{sec:numtests} were observed to be satisfactory in this respect. As an example,
TC tests with $\delta \Sigma_3/P= 5\times 10^{-3}$ and $\delta\Sigma_3/P = 2\times 10^{-2}$ yield consistent results.

Finally, finite-size effects are expected~\cite{PR08a}, hence a fifth dimensionless parameter in the problem, the number $N$ of particles.
Values of the dimensionless control parameters in the presently reported simulations are listed in Tab.~\ref{tab:param}.
\begin{table}
\begin{ruledtabular}
\begin{tabular}{ccccc}
$\kappa$ & $I$ & $\zeta$ & $\delta\Sigma / P$ & $N$ \\ \hline
\{$8.4\times 10^{3}$,$3.9\times 10^4$\} & $< 10^{-4}$ & 0.98 & 0.005 & \{1372,4000,8788\}
\end{tabular}
\caption{Values taken by the dimensionless parameters.\label{tab:param}}
\end{ruledtabular}
\end{table}
We are chiefly interested in the \emph{macroscopic geometric limit}, in which all mechanical properties
are expected to depend on packing geometry alone, as announced in the introduction. It was defined in~\cite{PR08a} as the 
triple limit of $\kappa \rightarrow +\infty$ (rigid limit), $I \rightarrow 0$ (quasistatic limit)  and $N \rightarrow +\infty$ (thermodynamic limit). 
This limit was shown in Ref.~\cite{PR08a} to be correctly approached with the range of parameters displayed in Tab.~\ref{tab:param}. 

\section{Failure\label{sec:failure}}
The material being initially assembled in an isotropic state, the range of stress tensors it will sustain in the solid state can be defined in
principal stress space. From the known behavior of cohesionless granular materials (with friction in the contacts)~\cite{SAL77,MIT93,TH00,SUFL04} it is 
expected -- and it was explicitly checked in the case of S tests~\cite{PR08a} -- that the boundary of the set of supported stresses is reached on increasing
the deviatoric part of  $\ww{\sigma}$, away from the isotropic state. It is customary to define a loading function (or yield function) $f$ of 
 principal stresses $(\sigma_1,\sigma_2,\sigma_3)$, such that $f<0$ defines the region of supported stresses (which is believed to be convex in general~\cite{SAL77})
and  $f=0$ its boundary surface. 

For an assembly of perfectly rigid ($\kappa = +\infty$), noncohesive grains, 
the absence of stress scale implies that $f(\lambda\,\sig) = f(\sig)$ for all $\lambda > 0$.  
Thus, the failure surface of such a material has a conical shape in principal stress space. We assume this property 
to hold for our simulated system, which is close to the rigid limit. 
Consequently, it is sufficient to determine the  intersection of the failure surface with one deviatoric plane, 
that is the failure curve. 
Furthermore, because of the isotropic preparation method employed, $f$ is a symmetric function of $(\sigma_1,\sigma_2,\sigma_3)$
and the failure curve is left invariant by all permutations of $(\sigma_1,\sigma_2,\sigma_3)$.

The (cohesionless) Mohr-Coulomb model
\[
f_{\rm MC}(\ww{\sigma}) = \sigI - \sigIII - (\sigI + \sigIII) \sin\varphi
\]
is often assumed (at least implicitly) true for granular materials~\cite{Ned92}. 
Numerous studies have been devoted to the macroscopic friction of sheared granular assemblies in various geometries~\cite{Gdr04}, 
and it is tempting to assume that the  measured angle corresponds to friction angle $\varphi$ in a 
Mohr-Coulomb model that would describe the failure properties of the material. 
The fourth column of Tab.~\ref{tab:model} displays the macroscopic friction angles measured with the three loading paths employed. 
It shows that $\varphi$ depends on $N$, as already observed in Ref.~\cite{PR08a}, but also on the kind of mechanical test employed. 
Consequently, the material cannot be described by a Mohr-Coulomb criterion. 
Although TC and TE tests are not sufficient to rule out the Mohr-Coulomb model since 
$\varphi^{\rm TE} - \varphi^{\rm TC}$ is below the statistical uncertainties vitiating the results, 
the comparison with the \testtrois{} angles unambiguously invalidates this criterion.

\begin{table}
\begin{ruledtabular}
\begin{tabular}{ccccccc}
Test & $N$ & $S_N$ & $\varphi$ & $\Delta \varphi$ & $k$ & $\Delta k$ \\ \hline
\rule{0mm}{4.5mm}\multirow{3}{*}{TC} & 1372 & 8 & $8.3^{\circ}$ & $0.6^{\circ}$ & $27.80$ & $0.11$ \\
                    & 4000 & 8 & $6.8^{\circ}$ & $0.5^{\circ}$ & $27.52$ & $0.08$ \\
                    & 8788 & 8 & $6.0^{\circ}$ & $0.2^{\circ}$ & $27.41$ & $0.03$ \\
\hline
\rule{0mm}{4.5mm}\multirow{3}{*}{TE} & 1372 & 8 & $8.6^{\circ}$ & $0.5^{\circ}$ & $27.81$ & $0.10$ \\
                    & 4000 & 8 & $6.9^{\circ}$ & $0.2^{\circ}$ & $27.52$ & $0.02$ \\
                    & 8788 & 8 & $6.0^{\circ}$ & $0.4^{\circ}$ & $27.40$ & $0.05$ \\
\hline
\rule{0mm}{4.5mm}\multirow{3}{*}{S}  & 1372 & 6  & $9.7^{\circ}$ & $0.3^{\circ}$ & $27.80$ & $0.04$ \\
                    & 4000 & 10 & $7.8^{\circ}$ & $0.3^{\circ}$ & $27.51$ & $0.07$ \\
                    & 8788 & 6  & $7.0^{\circ}$ & $0.4^{\circ}$ & $27.41$ & $0.04$ \\
\end{tabular}
\caption{Macroscopic friction angle $\varphi$ and Lade-Duncan parameter $k$, 
measured just before failure on $S_N$ distinct initial configurations, for different mechanical tests and different system sizes. $\Delta \varphi$ and 
$\Delta k$ are the corresponding standard deviations.
\label{tab:model}}
\end{ruledtabular}
\end{table}

It would be appealing if one could characterize the failure properties of the material with a single parameter
that would not depend on the applied load direction. 
Such an attempt already proved successful for assemblies of frictional equal-sized beads~\cite{TH00,SUFL04}, 
whose failure curve was successfully modeled by a Lade-Duncan criterion~\cite{LD75}:
\be
f_{\rm LD}(\ww{\sigma}) = \frac{(\sigI + \sigII + \sigIII)^3}{\sigI\sigII\sigIII} - k
\label{eq:lade}
\ee
One should have $k\ge 27$ in~\eqref{eq:lade} if condition $f\le 0$ is to define a non-empty cone of supported stresses (with $k=27$ only
isotropic stresses would be possible in equilibrium and the material would behave like a liquid).
According to Tab.~\ref{tab:model}, this failure criterion also works well in the frictionless case: the values of $k$ deduced from  
each of the three loading paths (see Tab.~\ref{tab:critparam}) agree with one another.
\begin{table}
\begin{ruledtabular}
\begin{tabular}{ccc}
Test & $\sin\varphi$ & k \\ \hline
TC   & $\dfrac{\Sigma_3 - \Sigma_1}{\Sigma_3 + \Sigma_1}$ & $\dfrac{(3 - \sin\varphi)^3}{1 - \sin\varphi - \sin^2\varphi + \sin^3\varphi}$ \\
TE   & $\dfrac{\Sigma_1 - \Sigma_3}{\Sigma_1 + \Sigma_3}$ & $\dfrac{(3 + \sin\varphi)^3}{1 + \sin\varphi - \sin^2\varphi - \sin^3\varphi}$ \\
S    & $\dfrac{\tau}{P}$                                  & $\dfrac{27}{1-\sin^2\varphi}$
\end{tabular}
\caption{Mohr-Coulomb and Lade-Duncan parameters for the different loading paths, as functions of applied stress components.
 $\sin\varphi$, as defined in the second column, is used as an intermediate variable in the expression of parameter $k$ in the third one. 
\label{tab:critparam}}
\end{ruledtabular}
\end{table}

\begin{figure}
\centering
\includegraphics[height=6cm]{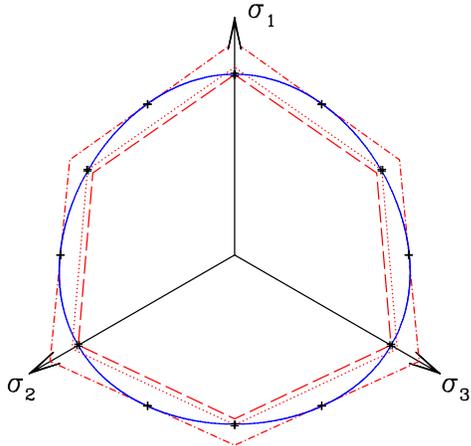}
\caption{(Color online) Calculated points with their error bars, Lade-Duncan criterion ($k = 27.80$, blue solid line) 
and cohesionless Mohr-Coulomb criteria corresponding to TC ($\varphi = 8.3^{\circ}$, red dashed line), 
TE ($\varphi = 8.6^{\circ}$, red dotted line), and S tests ($\varphi = 9.7^{\circ}$, red dotted and dashed line)
 for an assembly of $N = 1372$ particles.\label{fig:ldVSmc}}
\end{figure}

Making use of the aforementioned permutation symmetry, the stresses at failure
computed for $N=1372$ are plotted in Fig.~\ref{fig:ldVSmc}, in the deviatoric plane, 
with the Lade-Duncan curve corresponding to $k(1372)$ and the three Mohr-Coulomb failure curves pertaining to the three distinct numerical tests performed. 
The Lade-Duncan criterion is clearly the best model.

\begin{figure}
\centering
\includegraphics[height=6cm]{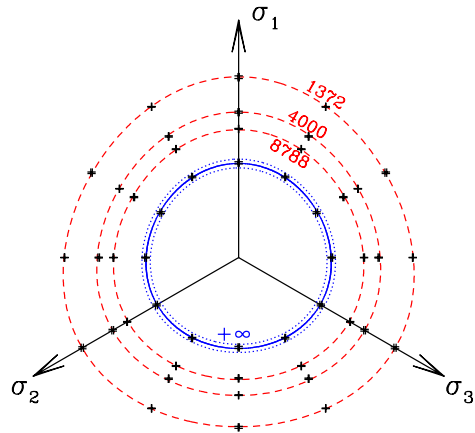}
\caption{(Color online) Principal stresses at failure  and corresponding Lade-Duncan fits for $N = 1372$, $4000$ and $8788$ (red dashed lines). 
The blue solid curve corresponds to the macroscopic limit (k = 27.22), whereas the two blue dotted curves, for $k = 27.20$ and $k = 27.24$,
bound the uncertainty interval.\label{fig:ldN}}
\end{figure}

Failure properties are, however, dependent on system size (Tab.~\ref{tab:model}). 
Fig.~\ref{fig:ldN} plots the principal stresses at failure in the deviatoric plane for $N = 1372$, $4000$, $8788$ with the corresponding Lade-Duncan fits. 
The domain bounded by the failure limit decreases with increasing $N$. 
To evaluate the failure curve in the macroscopic limit of $N\to+\infty$, principal stresses obtained in finite-size samples are extrapolated, 
assuming a linear dependence with $N^{-1/2}$. This assumption is proved to be statistically valid thanks to $\chi^2$ calculations~\cite{numrec}
and the resulting principal stresses are plotted in Fig.~\ref{fig:ldN}.
A Lade-Duncan fit of these extrapolated points with parameter $k_{\infty} = 27.22 \pm 0.02$ works well. 
As $k_{\infty} > 27$, the failure surface is not reduced to the trisectrix in the $N \to +\infty$ limit, and  
macroscopic systems can be equilibrated under moderately anisotropic loads, in agreement with~\cite{PR08a}.
The value taken by the Lade-Duncan parameter in the $N \to +\infty$ limit corresponds to 
$\varphi_{\infty}^{\rm TC} = 4.4^{\circ} \pm 0.2^{\circ}$ in triaxial compression, 
$\varphi_{\infty}^{\rm TE} = 4.5^{\circ} \pm 0.3^{\circ}$ in triaxial extension, and $\varphi_{\infty}^{\rm S} = 5.2^{\circ} \pm 0.3^{\circ}$ for \testtrois{} tests. 
The latter value agrees with the static friction angle given in Ref.~\cite{PR08a} in the macroscopic geometric limit.

One can observe that the shape of the criterion in the deviatoric plane becomes more rounded with increasing $N$. 
Since other criteria predict a nearly circular failure curve for small deviatoric strength~\cite{Maio06},
other forms than the Lade-Duncan yield function could be fitted to the
data in the macroscopic limit. First of all, we observed that the Dr\"ucker-Prager criterion~\cite{DP52} 
(whose shape is always circular in the deviatoric plane) does not correctly fit our results in the $N \to +\infty$ limit. 
The Matsuoka-Nakai criterion~\cite{MN85}, a model specifically tailored to capture the failure properties of some sands, defined as
\[
f_{\rm MN}(\ww{\sigma}) = \frac{(\sigI + \sigII + \sigIII)(\sigI \sigII + \sigII \sigIII + \sigIII \sigI)}{(\sigI \sigII \sigIII)} - m,
\]
accurately fits the data extrapolated in the macroscopic limit (with $m = 9.05 \pm 0.01$). 
Note however that this criterion is not suitable to describe failure in the smallest finite-size systems studied.

In general, the expressions of failure criteria (Lade-Duncan, Matsuoka-Nakai, or Mohr-Coulomb) are purely phenomenological, and their justification is 
to provide a convenient fit function. In the present case, stress anisotropies will be related to other internal variables in Sec.~\ref{sec:before}, but
a prediction of the shape of the failure curve in the deviatoric plane 
(related to complex geometric properties of sphere packings) is currently beyond our reach.

\section{Solid behavior and microstructure: the road to failure\label{sec:before}}
We now study the evolution of the material within the solid range, from the initial isotropic state to the failure limit,
with a particular emphasis on microscopic aspects.
We first investigate in Sec.~\ref{sec:scal} how the scalar variables characterizing the internal state of the packing evolve with growing deviator stress. 
Those variables include solid fraction $\Phi$, connectivity and coordination number, orientation-averaged pair correlations and force distributions,
and were extensively studied in isotropic RCP states~\cite{DTS05,iviso1}. 
Structural and force anisotropies~\cite{RWJM98,KR96,BR88,RB89} are studied in Sec.~\ref{sec:anisostruc}.  
In the spirit of~\cite{ARPS07}, we will show how stresses relate to anisotropy parameters.
Then, Section~\ref{sec:elastic} reports on the elastic moduli measured in nearly rigid anisotropic packings,
with results generalizing previous numerical observations on isotropic RCP state elastic properties.
Finally, the existence of a well-defined stress-strain law in the thermodynamic limit ($N \rightarrow +\infty$) is discussed in Sec.~\ref{sec:thermolim}.

\subsection{Scalar quantities\label{sec:scal}}
\begin{figure}
\includegraphics[width=8.5cm]{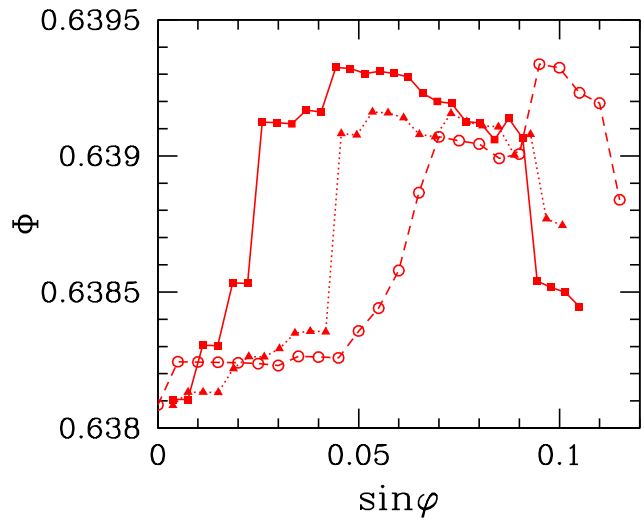}
\caption{(Color online) Volume fraction $\Phi$ as a function of $\sin\varphi$ (as defined in Table~\ref{tab:critparam})
for $N = 8788$ and $\kappa = 3.9\times 10^4$.
Solid squares are for one TC test, solid triangles for one TE test and circles for one S test. 
Curves are stopped at the value of $\sin\varphi$ corresponding to the failure limit.\label{fig:volfrac}}
\end{figure}

The typical evolution of volume fraction $\Phi$ with the deviatoric stress applied, 
characterized by $\sin \varphi \equiv (\sigI - \sigIII) / (\sigI + \sigIII)$, is depicted in Fig~\ref{fig:volfrac}. 
It shows that whatever the load applied, $\Phi$ remains approximately equal to $\Phi_{\rm RCP} \simeq 0.639$
from the initial isotropic state to the failure threshold. In particular, the relative variations of $\Phi$ remain smaller by more
than an order of magnitude than the deviatoric strains (see Sec.~\ref{sec:thermolim}). 
This is consistent with Ref.~\cite{PR08a} which showed the material to be devoid of dilatancy in the macroscopic geometric limit.
$\Phi$ evolves quite erratically with $\sin\varphi$; however, $\Phi$ seems to increase systematically when the applied stress is moderately anisotropic,
then it reaches a maximum and finally, it decreases when the material approaches its failure limit. We have currently no convincing explanation for this phenomenon.
The jumps in $\Phi$ are correlated to network rearrangements: we checked that the greater the jump, the more important the change in the contact list. 

The connectivity of the contact network is the set ($p_n$) of probabilities for one grain to be involved in $n$ contact forces. 
The coordination number $z$ is linked to ($p_n$) through $z = \sum_n np_n$. 
Average fractions $p_n$ have been recorded for the three loading paths with $N = 1372$, $4000$ and $8788$. 
At equilibrium, whatever the deviator applied, the set ($p_n$) is found 
identical to the distribution measured on frictionless isotropic packings~\cite{iviso1}. For such packings, $p_1$, $p_2$ and $p_3$ vanish, 
because normal repulsive forces on a bead with less than four contacts cannot balance. 
As in the case of isotropic packings, some grains, the rattlers, do not belong to the force-carrying structure: 
their proportion is estimated at $p_0 \simeq 1.3\%$, which is close to the value obtained with isotropic packings~\cite{iviso1}. 
In all simulations carried out with $\kappa = 3.9\times 10^4$, the backbone coordination number $z^* = z(1-p_0)^{-1}$ remains equal to $6.08 \pm 0.03$ 
between the initial isotropic state and the failure limit. 
By the isostaticity property of the backbone, $z^*$ tends toward $6$ in the $\kappa \to +\infty$ limit~\cite{OSLN03,iviso1,PR08a}.

\begin{figure}
\includegraphics[width = 8.5cm]{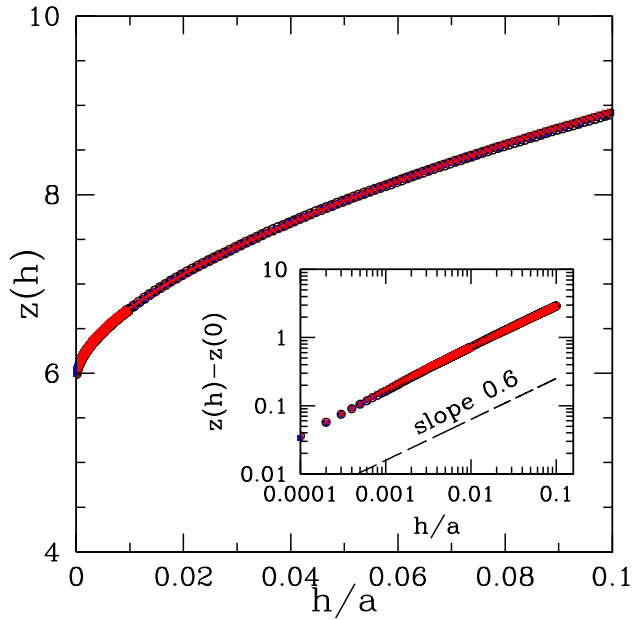}
\caption{(Color online) Average coordination number $z(h)$  of network $\mathcal{C}_h$
as a function of $h/a$, computed from some equilibrated configurations near their failure limit under TC, TE and S tests.
All three group of data collapse on a single curve. Inset: power law behavior of $z(h)-z(0)$ for $h/a \ll 1$, revealed by a double logarithmic plot.\label{fig:zh}}
\end{figure}

If we now replace the contact network by network $\mathcal{C}_h$ defined on declaring a bond to join all pairs of grains
separated by a distance smaller than $h$, then its coordination number $z(h)$ is drawn as a function of $h/a$ in Fig.~\ref{fig:zh}
at the failure limit. Curves corresponding to the three studied loading paths are identical. 
$z(h)$ starts from coordination number $z$ at $h=0$ and is the cumulated integral of the pair correlation function up to distance $a+h$
between sphere centers. One gets $z(h) - z(0) \propto (h/a)^{0.6}$ for $h/a \ll 1$ in all equilibrated packings.
The same power law with exponent $0.6$ has already been observed to fit $z(h)$
data in the same range of gap $h$ with isotropic packings (RCP states)~\cite{DTS05,iviso1}. 
(No theoretical basis has been proposed for this power law, 
the prefactor and the exponent of which might slightly depend on the range of $h$ fitted and on the treatment of rattlers~\cite{SiLiNa06,iviso1}.)

The probability distribution functions $p(f)$ of normalized contact forces, $f=F/\ave{F}$, has a similar shape as reported
in many numerical~\cite{RJMR96,OR97b,SGL02,DTS05,iviso1} and some experimental~\cite{CLMNW96,BMMJN01} studies on granular media. 
$p(f)$ first exhibits a slight increase, up to $f\simeq 0.5$, and then it decreases, roughly exponentially for large $f$.
Remarkably, thanks to Kolmogorov-Smirnov tests~\cite{numrec}, we observed that all p.d.f. in the equilibrium configurations obtained for the different simulated stress states coincide. 
Neither the number of grains, nor the direction of the loading path, nor the proximity of the failure limit alter the force distributions,
which remain statistically indistinguishable. $p(f)$ thus coincides, within statistical uncertainties, with the form parametrized, e.g., in~\cite{DTS05}. 
As the backbone is isostatic, $p(f)$ is geometrically determined in the rigid limit.
The p.d.f. may in particular be characterized by its moments, which we denote, for any $x>0$, as
\be
Z(x) = \ave{f^x}=\frac{\ave{F^x}}{\ave{F}^x},
\label{eq:defZ}
\ee
and we obtain, \emph{e. g.}, $Z(2)=1.53\pm 0.02$ and $Z(5/3) = 1.29\pm 0.01$ (those results will be useful in Sec.~\ref{sec:elastic}).

Finally, using the same indicators as in~\cite{iviso1}, we observed no tendency towards the formation of locally crystalline patterns in the configurations
under varying deviator stresses.
\subsection{Anisotropy\label{sec:anisostruc}}
Previous works showed that the very origin of shear strength in granular materials is the anisotropy, both structural and mechanical,
induced by the deviatoric stress~\cite{R08}. We now explore this connection in the particular case of frictionless, rigid bead assemblies.

Mathematically, material anisotropy can be characterized by the joint probability density function $P(\vec{n},F)$ of
finding an intergranular contact oriented along the unit vector $\vec{n}$ and carrying a force of intensity $F$.
This quantity is of central importance since it intervenes in the expression of the Cauchy stress tensor.
Bearing in mind that $\kappa \gg 1$ and denoting the number of contacts by $N_c$, Eq.~\eqref{eq:stresstensor} can be rewritten as:
\bea
\ww{\sigma} & = & \frac{N_c a}{V} \ave{\vec{F} \otimes \vec{n}} \nonumber \\
            & = & \frac{N_c a}{V} \int \textrm{d}\Omega\,\textrm{d}F\, P(\vec{n},F) F \vec{n} \otimes \vec{n} \nonumber \\
            & = & \frac{N_c a}{V} \int \textrm{d}\Omega\,E(\vec{n})\ave{F}_{\vec{n}} \:\vec{n} \otimes \vec{n} \label{eq:sigma}
\eea
$\ave{F}_{\vec{n}}$ is the angular force density (it is equal to $\ave{F}/(4\pi)$ in the isotropic case) and $E(\vec{n})$
is the probability density function of finding a contact along $\vec{n}$.

\subsubsection{Structural anisotropy and its relation to stress ratios\label{sec:fabric}}
The anisotropy of the contact network is described by $E(\vec{n})$. $E$ is defined on the unit sphere of $\mathbb{R}^3$,
so it can be expanded in a series of spherical harmonics. Since contacts are undirected, odd order coefficients in the expansion vanish. 
At the lowest order, the expansion is restricted to the spherical harmonics of order $2$ and the coefficients are related to the second-order fabric tensor
$\ww{F} \equiv \ave{\vec{n} \otimes \vec{n}}$. Furthermore, for \testtrois{} tests,
it was shown in Ref.~\cite{PR08a} that a single anisotropic term of the expansion dominates:
\be
E(\vec{n}) \simeq \frac{1}{4\pi} + F_{12}\, d_{xy}(\theta,\psi)
\label{eq:Ecis}
\ee
with $d_{xy}(\theta,\psi) = 15\sin^2\theta \sin (2\psi) / (8\pi)$ 
($\theta$ is the colatitude angle and $\psi$ the longitude angle of the spherical coordinates). 
In the case of a triaxial test, by axial symmetry, the expansion of $E$ in spherical harmonics up to the second order reads:
\be
E(\vec{n}) \simeq \frac{1}{4\pi} + (F_{33}-\frac{1}{3}) d_{z^2}(\theta,\psi)
\label{eq:Etriax}
\ee
with $d_{z^2} = 15(3\cos^2\theta - 1) / (16\pi)$.
\begin{figure}
\includegraphics[width=8.5cm]{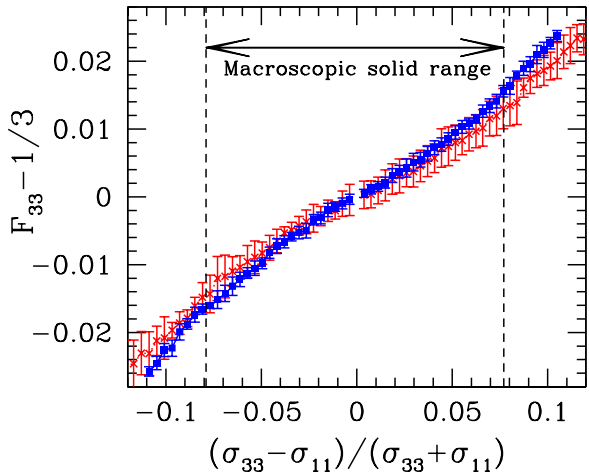}
\caption{(Color online) Average evolution of the structural anisotropic term with $r = (\sigma_{33}-\sigma_{11})/(\sigma_{33}+\sigma_{11})$
under TC ($r > 0$) and TE ($r < 0$) tests. Red crosses correspond to $N = 1372$ and blue squares to $N = 8788$.\label{fig:fabric}}
\end{figure}

Fig.~\ref{fig:fabric} shows how the anisotropic term evolves with $(\sigma_{33}-\sigma_{11})/(\sigma_{33}+\sigma_{11})$ ($\sigma_{11}$ and $\sigma_{33}$ are 
principal stresses under a triaxial load), for systems of two different sizes subjected to a TC or TE test. 
Whatever the test performed, the absolute value of the anisotropic term increases with the applied deviator intensity. 
An analysis of the regression of fluctuations for the data of Fig.~\ref{fig:fabric} indicates that the evolution of the anisotropic
terms with stress deviator intensity tends to a well-defined curve in the macroscopic limit.
The dependence is roughly linear, even if one can notice that the slope of the curves seems
to change around the boundaries of the solid range in the limit of $N \to +\infty$. (Other expressions involving principal stress ratios could have been
used to characterize stress anisotropy).
Although the maximum value of the anisotropy parameter is size-dependent, 
the slope $\sfab$ of the straight line fitting in the macroscopic solid range the (sample-averaged) anisotropy parameter
as a function of $(\sigma_{33}-\sigma_{11})/(\sigma_{33}+\sigma_{11})$ for triaxial tests, and of $\sigma_{12}/\sigma_{22}$
for \testtrois{} tests, does not depend on $N$ if the number of grains is large enough. For $N \geq 4000$,
numerical simulation yield $\sfab = 0.197 \pm 0.010$ for TC tests,
$\sfab = 0.210\pm 0.015$ for TE tests, and $\sfab = 0.158 \pm 0.015$ for S tests. 

\begin{figure}
\includegraphics[width=8.5cm]{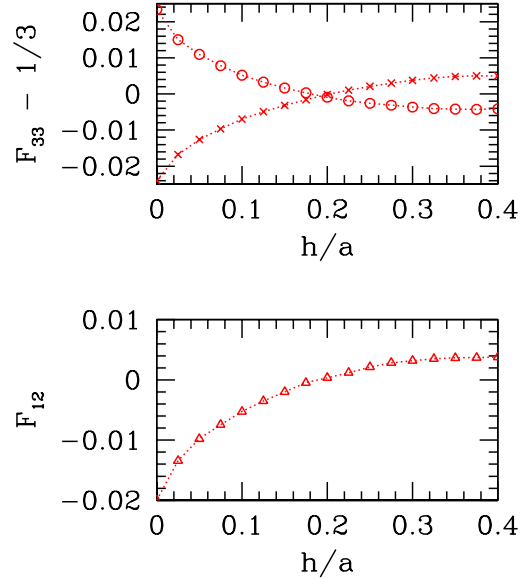}
\caption{(Color online) Dominant structural anisotropic term at the failure limit as a function of the gap $h$ 
ith $N = 8788$ for TC tests (circles), TE tests (crosses) and S tests (triangles).\label{fig:fabrich}}
\end{figure}

The range of anisotropic pair correlations can be studied by considering the fabric tensor of network $\mathcal{C}_h$
(defined in Section~\ref{sec:scal}) as a function of $h$. Anisotropy parameters are plotted
as functions of $h$ in Fig.~\ref{fig:fabrich}, for maximum stress anisotropies (at the failure limit). 
They first decrease for increasing $h$, and reach zero near $h/a=0.2$. The small values of opposite sign measured at larger distances
are of the order of the statistical noise ($\simeq 0.001$) observed on isotropic configurations and should be interpreted with care. The spatial distribution of near, but distant neighbors thus tends to cancel the anisotropy of the distribution of contacting ones. The material anisotropy is short-ranged. In particular it is very nearly negligible on averaging over the complete first neighbor shell (\emph{i. e.} up to the distance corresponding to the first minimum in the pair correlation function, $h/a\simeq 0.35$ from Refs.~\cite{DTS05,iviso1}).

\subsubsection{Force anisotropy and its relation to stress ratios\label{sec:fanis}}
The mechanical anisotropy is described by the angular dependence of $\ave{F}_{\vec{n}}$.
Like $E(\vec{n})$, it can be expanded in a series of spherical harmonics of even order.
To make things easier, only the expansion up to the second order is considered. 
As for the structural anisotropy, a single term, with the same symmetry, was assumed to dominate. For \testtrois{} tests
\be
\ave{F}_{\vec{n}} \simeq \left( \frac{1}{4\pi} + H_{12}\, d_{xy}(\theta,\psi) \right) \ave{F}
\label{eq:Fcis}
\ee
and for triaxial tests
\be
\ave{F}_{\vec{n}} \simeq \left( \frac{1}{4\pi} + H_{33}\, d_{z^2}(\theta,\psi) \right) \ave{F}
\label{eq:Ftriax}
\ee
with $\ave{F}$ the average force intensity.

Force anisotropy parameters $H_{12}$ and $H_{33}$ are obtained by dividing the unit sphere in small regions.
This allows to compute some values of $\ave{F}_{\vec{n}}$ , and coefficients  $H_{12}$ and $H_{33}$ are then derived 
by calculating the scalar product---defined as $\ave{f,g} = \int (\textrm{d}\Omega / (4\pi)) f(\theta,\psi) g(\theta,\psi)$, 
with $f$ and $g$ two functions defined on the unit sphere---of $\ave{F}_{\vec{n}}$ with $d_{xy}$ and $d_{z^2}$.

\begin{figure}
\includegraphics[width=8.5cm]{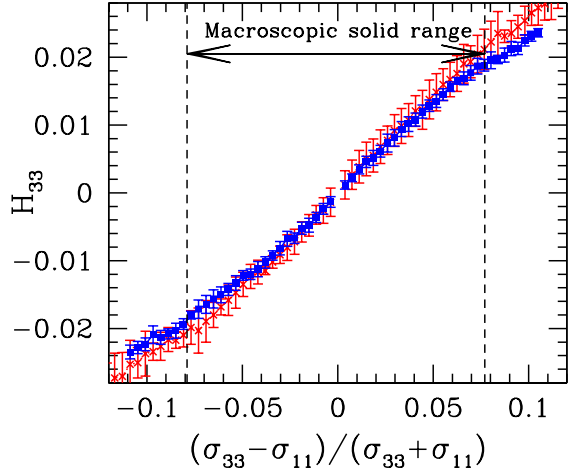}
\caption{(Color online) Average evolution of the mechanical anisotropic term with 
$r = (\sigma_{33}-\sigma_{11})/(\sigma_{33}+\sigma_{11})$ under TC ($r > 0$) and TE ($r < 0$) tests.
Red crosses correspond to $N = 1372$ and blue squares to $N = 8788$.\label{fig:mechani}}
\end{figure}

The build-up of $H_{33}$ under a triaxial load for two different system sizes is displayed on Fig.\ref{fig:mechani}.
It is very similar to the build-up of $F_{33}-1/3$. The numerical data evidence a one-to-one correspondence with stress anisotropy,
which is approximately linear for moderate deviators. The slope $\sfor$ of the plot of Fig.~\ref{fig:mechani}
seems to be independent of $N$ when $N$ is sufficiently large. With $N \geq 4000$,
one has $\sfor = 0.250 \pm 0.012$ for TC tests, $\sfor=0.235 \pm 0.015$ for TE tests, and $\sfor = 0.173 \pm 0.014$ for S tests (for which
$H_{12}$ relates to stresses approximately as $H_{12}= \sfor \dfrac{\sigma_{12}}{\sigma_{22}}$).
\subsubsection{General connection between stress and anisotropy\label{sec:connec}}
The observed relations between stress and fabric (Sec.~\ref{sec:fabric}) or force (Sec.~\ref{sec:fanis}) anisotropies were not, to our knowledge, previously
reported in the literature. We argue below in Section~\ref{sec:thermolim} that they are specific to frictionless grains in the rigid limit.

Yet, a more general connection between stress and both fabric and force anisotropies can be derived on using
spherical harmonics expansions for the relevant stress components, as deduced
from Eqs.~(\ref{eq:Etriax},\ref{eq:Ftriax}) for triaxial tests and from Eqs.~(\ref{eq:Ecis},\ref{eq:Fcis}) for \testtrois{} tests. Such a relation was
repeatedly used for frictional systems, most often in 2D~\cite{RB89,ARPS07,R08}.

In the case of triaxial tests, keeping only the terms up to the second order yields
\[
E(\vec{n}) \times \ave{F}_{\vec{n}} \simeq \left[ \frac{1}{16\pi^2} + \left( \frac{H_{33}+F_{33}-1/3}{4\pi} \right) d_{z^2}(\theta,\psi) \right] \ave{F}
\label{eq:prodtriax}
\]
Combining this relation with Eqs.~\eqref{eq:sigma}, one gets
\bea
\sigma_{11} & \simeq & \frac{N_c a \ave{F}}{V} \left[ \frac{1}{12\pi} - \frac{1}{8\pi} (H_{33} + F_{33} - 1/3) \right] \notag \\
\sigma_{33} & \simeq & \frac{N_c a \ave{F}}{V} \left[ \frac{1}{12\pi} + \frac{1}{4\pi} (H_{33} + F_{33} - 1/3) \right]\notag
\eea
Consequently, one obtains
\be
\frac{\sigma_{33}}{\sigma_{11}} \simeq 2 \frac{H_{33} + F_{33}}{1 - H_{33} - F_{33}}. \label{eq:strratiotriax}
\ee

\begin{figure}
\includegraphics[width=8.5cm]{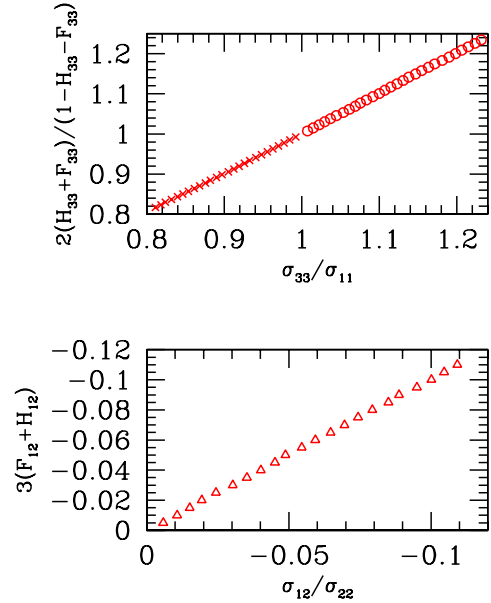}
\caption{(Color online) Numerical test of the approximations given by Eqs.~\eqref{eq:strratiotriax} and~\eqref{eq:strratiocis} with 
$N = 8788$ for TC tests (circles), TE tests (crosses) and S tests (triangles).\label{fig:stressani}}
\end{figure}

In the case of \testtrois{} tests, neglecting terms of order larger than $2$ yields
\[
E(\vec{n}) \times \ave{F}_{\vec{n}} \simeq \left[ \frac{1}{16\pi^2} + \left( \frac{F_{12}+H_{12}}{4\pi} \right) d_{xy}(\theta,\psi) \right] \ave{F} \label{eq:prodcis}
\]
By inserting the above equation in~\eqref{eq:sigma}, one gets
\bea
\sigma_{12} & \simeq & \frac{F_{12}+H_{12}}{4\pi} \frac{N_c a \ave{F}}{V},\notag  \\
\sigma_{22} & \simeq & \frac{1}{12\pi} \frac{N_c a \ave{F}}{V},\notag 
\eea
hence the result:
\be
\frac{\sigma_{12}}{\sigma_{22}} \simeq 3(F_{12}+H_{12}). \label{eq:strratiocis}
\ee
Although Eqs.~\eqref{eq:strratiotriax} and~\eqref{eq:strratiocis} are simple approximations, they work surprisingly well, as shown by Fig.~\ref{fig:stressani}.

In Sections~\ref{sec:fabric} and \ref{sec:fanis}, fabric and force anisotropies were separately related to stress ratio
(with approximate, linear relations involving parameters $\sfab$ and $\sfor$). Thus one should check for the
consistency between such observations and relations~\eqref{eq:strratiotriax} and \eqref{eq:strratiocis}. In the case of triaxial tests, on 
writing down all quantities to first order in the small anisotropy parameters $H_{33}$ and $F_{33}-1/3$, one obtains the consistence condition:
\be
\sfor +\sfab = \frac{4}{9}.
\label{eq:consistri}
\ee
Similarly, for S tests one should have:
\be
\sfor +\sfab = \frac{1}{3}.
\label{eq:consiscis}
\ee
The values of $\sfab$ and $\sfor$ obtained in Sections~\ref{sec:fabric} and \ref{sec:fanis} satisfy conditions~\eqref{eq:consistri} and \eqref{eq:consiscis} with 
good accuracy.

The simple connection between stress and anisotropy parameters expressed by Eqs.~\eqref{eq:strratiotriax} and \eqref{eq:strratiocis} 
emphasizes the microscopic origin of a macroscopic quantity
(a stress ratio in this case). In view of the different internal fabric symmetries in the triaxial and the shear tests, it is finally not surprising that the
corresponding friction angles differ (whence the inadequacy of the Mohr-Coulomb criterion, Section~\ref{sec:failure}).

In granular materials with friction, 
the shape of the particles influences the relative roles of geometry and mechanics in the sustained stress~\cite{ARPS07}.
For frictionless spherical grains, near the failure limit, 
we find that the parameters describing both anisotropies are approximately equal, so that half of stress ratio $\sigma_{33}/\sigma_{11}$ or $\sigma_{12}/\sigma_{22}$ is explained by geometric anisotropy and the other half by mechanical anisotropy.

Despite those simple relations between stresses and anisotropy parameters, 
theoretically predicting the stress ratio at failure still remains a challenge.

\subsection{Elastic moduli\label{sec:elastic}}
The motivation for computing elastic moduli is twofold. First, 
elastic properties are usually more easily measured in the laboratory than geometric data such as near
neighbor correlations and coordination numbers, as discussed in
Ref.~\cite{iviso3}. Then, the elastic moduli of frictionless bead packs under isotropic stresses were studied 
by numerical simulations~\cite{OSLN03,iviso3}, and shown to
exhibit singular properties, which we now seek to generalize to anisotropic stress states. Specifically, while the
bulk modulus, $B$, shows little difference with well coordinated frictional packings~\cite{iviso3},
the shear modulus, $G$, is anomalously small. 
$G/B$ tends to vary proportionally to the degree of force indeterminacy~\cite{Wyart-th2,iviso3}, which
vanishes in the rigid limit, as $\kappa^{-1/2}$. Isotropic frictionless bead packs also possess stiffness matrices (or ``dynamical matrices'') with 
an anomalous distribution of eigenmode frequencies~\cite{OSLN03}, which stems from the nearly isostatic character of the contact network~\cite{WNW05}.

\begin{figure}
\includegraphics[angle=270,width=8.5cm]{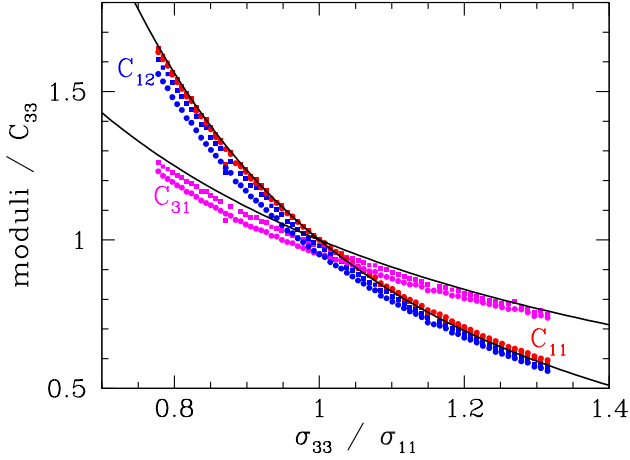}
\caption{
\label{fig:modrapp}
(Color online) Ratios $C_{11}/C_{33}$ (red),  $C_{13}/C_{33}$ (pink), and $C_{12}/C_{33}$ (blue), in TE and TC tests, versus
principal stress ratio, for Hertzian (square dots)
and linear (round dots) contact elasticity and $N = 8788$. Continuous (black) lines correspond to the predictions of Eqs.~\eqref{eq:appratio}.}
\end{figure}

For simplicity, we restrict our investigations to the elastic moduli of equilibrium configurations obtained in TC or TE tests.
They are numerically evaluated on building
the stiffness matrix of contact networks and solving linear systems of equations for displacements in response to small load increments,
as explained in~\cite{iviso3}.  The results are devoid of size effects and sample to sample fluctuations regress as $N$ increases.
There are five independent elastic constants in such cases (a number which would increase to nine for simple shear tests), which express a linear relation 
between stress increments $\Delta\sigma_{ij}$  and strains $\epsilon_{ij}$, from a reference equilibrium anisotropic state, as
\be
\bma \Delta\sigma_{11} \\ \Delta\sigma_{22} \\ \Delta\sigma _{33}  \\ \Delta\sigma_{23} \\ \Delta\sigma _{31} \\ \Delta\sigma_{12}\ema=
\bma C_{11} & C_{12} & C_{13} & 0 & 0 &0\\ C_{12} & C_{11} & C_{13} & 0 & 0 &0\\C_{13} & C_{13} & C_{33} & 0 & 0 &0\\
0&0&0&2C_{44}&0&0\\0&0&0&0&2C_{44}&0\\0&0&0&0&0&2C_{55}\ema 
\bma \epsilon_{11} \\ \epsilon_{22} \\ \epsilon _{33} \\ \epsilon_{23} \\ \epsilon_{31} \\ \epsilon _{12}\ema
\label{eq:modulanis}
\ee
The material symmetries -- invariance by rotation around axis 3 and by symmetry about all three planes of
coordinates -- determine the form of the matrix of elastic moduli
in~\eqref{eq:modulanis}, and also request that $C_{11}-C_{12}=2C_{55}$ (expressing the equality of two shear moduli in plane 1,2). Such symmetries are very well
satisfied (one has, e.g., $C_{13}=C_{23}$ with relative errors smaller than $10^{-3}$ for $N=8788$). 
The moduli in the initial isotropic state all relate to $B$ and $G$ as $C_{11}=B+4G/3=C_{33}$, $C_{12}=B-2G/3=C_{13}$, $C_{44}=C_{55}=G$. 
Then, longitudinal moduli (i.e., $C_{ii}$, with $i$=1, 2, 3) are larger in the direction of the major principal stress: thus one observes $C_{33}>C_{11}$
in triaxial compression and the opposite inequality in triaxial extension. This corresponds to different longitudinal sound wave velocities
$\sqrt{C_{ii}/\rho_m}$ ($\rho_m$ denoting the mass density of the material) propagating in direction 3 and in the orthogonal plane.
Such anisotropies of the elastic moduli were reported in the literature on sands~\cite{HaBl89,Tat195,DDBPE07} and bead packings~\cite{KhJi05}. They
can be attributed to the effect of both anisotropies, of fabric and forces, evidenced in Section~\ref{sec:anisostruc}: the material is stiffer in the
principal stress direction because it is favored in the distribution of contact orientations, and also because contacts nearly parallel 
to this direction tend to carry larger forces. As Hertz's law, Eq.~\eqref{eq:elas}, entails that $K_N\propto F_N^{1/3}$, such contacts are stiffer.  
To sort out the possible effects of fabric and force anisotropies,
we computed elastic moduli both for the Hertzian contact model and for linear contact elasticity, with
some constant, force-independent contact stiffness $K_N$. (As the packing geometry is very nearly that of a set of rigid beads, statistically 
similar configurations would have been obtained on simulating bead assemblies with linear
unilateral elastic contact forces). We focus in the sequel on the upper left
square block of order 3 within the matrix of moduli written in Eq.~\eqref{eq:modulanis}, which we denote as
$\ww{c}$. All of its elements are larger by about 2 orders of magnitude than
shear moduli $C_{44}$ and $C_{55}$, whatever the stress anisotropy. The ratio of all other elements of matrix $\ww{c}$ to $C_{33}$ are plotted
in Fig.~\ref{fig:modrapp}, for Hertzian and for linear contact elasticity.
The variations of $C_{11}/C_{33}$ with $\sigma_{33}/\sigma_{11}$ shown on Fig.~\ref{fig:modrapp} are qualitatively expected. More surprisingly, since 
the moduli evaluated with linear contact elasticity are not sensitive to force anisotropy, the dependence of such ratios on stress anisotropy
is about the same for both contact laws.

Such results, as we now explain, are due to the peculiar nature of matrix $\ww{c}$. Let $\hat s_1$, $\hat s_2$, $\hat s_3$ denote unit vectors in the space of
stress or strain tensors with eigendirections parallel to the coordinate directions, forming an orthonormal basis in which coordinates
are $\Delta\sigma_{ii}$ ($i$=1, 2, 3) (or $\epsilon_{ii}$) for stress (resp. strain) increments. Matrix $\ww{c}$ defines a linear operator within this space.
Under isotropic stresses, $\ww{c}$ has eigenvalues $C_I=3B$, $C_{II}=C_{III}=2G$, and eigenvectors are $\hat S_1 = (\hat s_1 +\hat s_2+\hat s_3)/\sqrt{3}$, and any pair
of vectors orthogonal to $\hat S_1$. The increment of stress in direction $\hat S_1$ is proportional to the preexisting equilibrium stress tensor (denoted here as
vector $P\sqrt{3}\hat S_1$). 
As an approximation, since $C_I \gg C_{II}$ and $C_I \gg C_{III}$, one may write:
\be
\ww{c}\simeq C_I \hat S_1 \otimes \hat S_1,
\label{eq:capprox}
\ee
bearing in mind that the right-hand-side is of course a singular matrix. On using~\eqref{eq:capprox}, all moduli would be equal to $B$ in the isotropic state and all
ratios equal to 1 in Fig.~\ref{fig:modrapp} for $\sigma_{11}=\sigma_{33}$, which is very nearly satisfied.
In \cite{iviso3}, it was argued that the ``dominant'' modulus,
$B$ is insensitive to the ``barely rigid'' character of the nearly isostatic contact network because it expresses the response to a load increment proportional 
to the preexisting load. We now apply similar ideas to anisotropic stress states. We first define $\hat S_1$ as the unit vector proportional to the preexisting,
equilibrium stress. The loading parameter in triaxial loading paths may be defined as $\alpha$ such that $\sigma_{11}=\sigma_{22}= (1-\alpha)P$
while $\sigma_{33}=(1+2\alpha)P$. We thus set:
\be
\hat S_1 = \frac{1}{\sqrt{3+6\alpha^2}}\left[ (1-\alpha)(\hat s_1 +\hat s_2) + (1+2\alpha)\hat s_3\right].
\label{eq:defE1}
\ee 
We observed $\hat S_1$ to be, with very good approximation, an eigenvector of $\ww{c}$, with eigenvalue $C_I$ close to its value in the isotropic state.
Due to the material symmetries in TC and TE tests, the second eigenvector should be $\hat S_2 = (\hat s_1 - \hat s_2)/\sqrt{2}$,
a property also well satisfied by the 
numerical data -- and the third one is of course orthogonal to  $\hat S_1$ and $\hat S_2$. We observed the corresponding eigenvalues $C_{II}$ and  $C_{III}$
to remain below $0.02\times C_I$ in all cases, whatever the stress anisotropy and the contact law (Hertzian or linear). Thus it is possible to approximate
matrix $\ww{c}$ on using relation~\eqref{eq:capprox}, with definition~\eqref{eq:defE1} for vector $\hat S_1$. This yields theoretical expressions for the
ratios between moduli:
\be
\frac{C_{11}}{C_{33}} \simeq \frac{(1-\alpha)^2}{(1+2\alpha)^2}\simeq \frac{C_{12}}{C_{33}} ;\ \ 
\frac{C_{13}}{C_{33}} \simeq \frac{1-\alpha}{1+2\alpha}. \label{eq:appratio}
\ee 
Fig.~\ref{fig:modrapp} shows that those approximations are quite accurate.
Thus stress anisotropies influence the tensor of elastic moduli in a peculiar way, due to its
nearly uniaxial, singular structure, which is independent of the contact law.
In a good approximation all moduli, except the very small, singular ones, are
proportional to $C_I$ with coefficients that are determined by the stress state.

\begin{figure}
\includegraphics[angle=270,width=8.5cm]{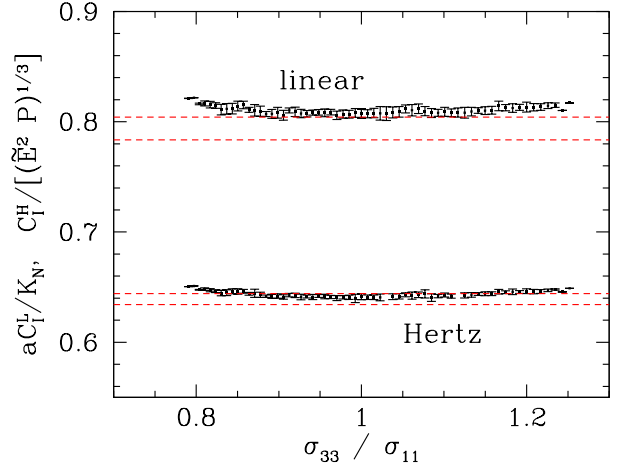}
\caption{
\label{fig:modred}
(Color online) Dominant eigenvalue $C_I$ of tensor of elastic moduli, for Hertzian and linear contact elasticity ($N=8788$), versus
principal stress ratio along TE and TC loading paths, compared to the predictions
of Eqs.~\eqref{eq:cIHertz} and~\eqref{eq:cIlin}, depicted, due to the slight statistical uncertainty, as narrow zones between horizontal dashed lines.
}
\end{figure}
On exploiting the isostaticity property of the contact network, it turns out that the dominant eigenvalue of tensor $\ww{c}$, $C_I$ 
can be written, in very good accuracy, as a simple function of solid fraction $\Phi$, 
coordination number $z$ and moments of the (geometrically determined) force distribution. Such a relation was established for the bulk
modulus $B$ of isotropic states in~\cite{iviso3}, where it is called the Reuss estimate. In general, it provides a lower bound to the modulus, which 
becomes exact when force increments are proportional to preexisting forces. This condition is exactly fulfilled by the response of
isostatic contact networks to an increment of stress tensor that is proportional to the preexisting stress tensor.
On adapting the approach followed in~\cite{iviso3} to the case of anisotropic stress states, one readily obtains, in the case of Hertzian contacts:
\be
C_I =C_I^H = \frac{3^{1/3}}{2Z\left(5/3\right)} \left( \frac{z\Phi}{\pi}\right)^{2/3} \tilde E^{2/3}P^{1/3}.
\label{eq:cIHertz}
\ee
For linear contact elasticity, the corresponding prediction reads
\be
C_I=C_I^L =\frac{z\Phi K_N}{\pi a Z(2)}.
\label{eq:cIlin}
\ee
$Z(5/3)$ and $Z(2)$ values are given after Eq.~\ref{eq:defZ}. All quantities appearing in those formulas were observed in 
Section~\ref{sec:scal} to remain constant throughout the range of supported stresses. Thus $C_I$ should not depend on principal stress ratio. 
Fig.~\ref{fig:modred} shows that the numerical data 
abide very well by the predictions of Eqs~\eqref{eq:cIHertz} and~\eqref{eq:cIlin}. Thus, all moduli, except
the soft ones that vanish in the rigid limit, are predicted. In the case of simple shear, we expect similar properties to apply, on adequately
redefining $\hat S_1$ in the direction of the applied load. In general, for arbitrary applied stresses within the supported range $f(\ww{\sigma})<0$ defined in 
Section~\ref{sec:failure}, one should have a nearly uniaxial tensor of elastic moduli. 

The stress increment or strain range for elastic response  
is expected to shrink to naught in the double limit of $\kappa\to+\infty$ and $N\to\infty$,
like the stability range of a contact network~\cite{CR2000}. Thus, in practice,
in order to observe the peculiar elastic properties of nearly rigid frictionless bead assemblies, one should adequately choose stiffness level $\kappa$, 
which should be large enough to approach the rigid limit but small enough for some elastic response to be measurable.
Interestingly, poorly coordinated packings of frictional disks~\cite{SRSvHvS05,SvHESvS07} or spheres~\cite{iviso3,MLRJWM08}
tend to exhibit similar elastic anomalies, although,
most often, in a weakened form, because such systems do not spontaneously form isostatic contact structures in the rigid limit~\cite{iviso1}. Even though truly
frictionless particles do not exist in the laboratory, our results might therefore bear some relevance in more general situations of contact networks
with quite a small level of force indeterminacy.
\subsection{Constitutive relations\label{sec:thermolim}}
\begin{figure}
\includegraphics[width=8.5cm]{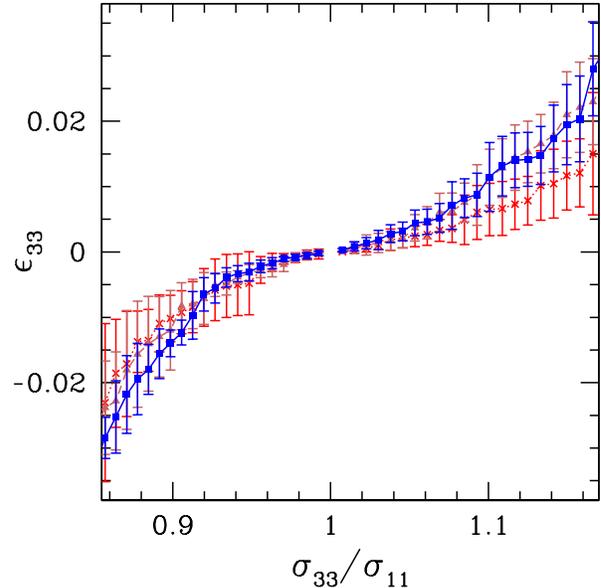}
\caption{(Color online) Evolution of strain $\epsilon_{33}$ with $\sigma_{33}/\sigma_{11}$ in triaxial tests, for $N = 1372$
(red crosses connected by a dotted line), $N = 4000$ (brown triangles connected by a dashed line),
and $N = 8788$ (blue squares connected by a solid line). Results are averaged over all available samples, and
restricted to the macroscopic solid range.\label{fig:defsize}}
\end{figure}
The observations of Sections~\ref{sec:anisostruc} and~\ref{sec:elastic} are strongly reminiscent of the results obtained on dealing with 
exactly rigid frictionless grains~\cite{CR2000,RC02}. If contacts are rigid, the response to an applied stress increment proportional to the
preexisting stress is also perfectly rigid: the corresponding strain is exactly zero. 
For all other stress increment orientations, the rigid contact network, in the limit of $N\to\infty$, has to rearrange~\cite{CR2000}.
The resulting strain is determined  by the geometry of the packing, rather than by some material stiffness.
Thus, using the notations of Sec.~\ref{sec:elastic}, $C_I$ is infinite, while 
$C_{II}=C_{III}=0$. 
This behavior also entails a one-to-one correspondence between stress and fabric anisotropy, in agreement
with Sec.~\ref{sec:fabric}. Because of isostaticity, the force distribution is completely determined by the force network, whence the
relation evidenced in Sec.~\ref{sec:fanis}. In this respect, assemblies of frictionless grains differ from systems with intergranular friction, in which one given
contact network may support stresses within a finite range in the thermodynamic limit, for arbitrary large stiffness levels $\kappa$~\cite{RC02} (whence
vertical parts in stress versus strain plots, as obtained in simulations with models of rigid grains~\cite{RTR04,RR04,R08}). This property of frictional
grain assemblies excludes the possibility of a one-to-one relation between stresses and fabric.

Rigid frictionless grain
assemblies, on the other hand, were reported~\cite{CR2000,RC02} 
to be devoid of the stress-strain relations (which depend on loading history) obtained in simulations of
model frictional systems~\cite{R08,TH00,SUFL04}, and classically modeled, for sands, in soil mechanics~\cite{DMWood,MIT93}. This conclusion was based on a statistical
analysis of the strain response to stress increments, which was modeled as a L\'evy-distributed random variable~\cite{BG90}, 
precluding the regression of strain fluctuations in the 
thermodynamic limit. Such results contrast with the ones obtained with particles interacting with soft potentials, such as Lennard-Jones glasses, in which case
fluctuations around the average stress-strain curve were explicitly shown to regress in the thermodynamic limit~\cite{MaLe06}.

In the present case, the macroscopic mechanical response is also dominated by packing rearrangements:
macroscopic strains are much larger than typical contact deflections
(of order $\kappa^{-1}$). Macroscopic strains, as plotted versus applied stress ratio along the triaxial test paths
in Fig.~\ref{fig:defsize}, do not appear to behave like a L\'evy flight trajectory: results pertaining to
the two larger sample sizes tend to cluster around the same average curve. However, the regression of fluctuations in the limit of $N\to\infty$ is much less
clearcut than in the results of, \emph{e.g.,} Fig.~\ref{fig:fabric}: error bars are only very slightly reduced between $N=1372$ and $N=8788$, and still extend to 
a notable fraction of averages (typically 30\%). Our 
data very likely provide unsufficient statistics because of sample size limitations, and larger systems should be studied. Yet it is tempting to speculate that
large enough samples, for given $\kappa$, do approach a well-defined stress-strain behavior for given loading paths, but that their size should exceed a certain 
characteristic length $\xi$ that diverges in the limit of $\kappa\to+\infty$. In this interpretation, for any given value of $\kappa$, samples of (linear)
size below $\xi$ would exhibit the singular behavior observed in Ref.~\cite{CR2000} (in which rigid contacts were simulated, with a specific numerical technique 
exploiting the isostaticity property). Only for samples larger than $\xi$ (and hence, for larger and larger samples as $\kappa$ is increased) should one recover
a well-defined stress-strain relationship in monotonic loading.
Further investigations of this conjecture are beyond the scope of the present paper.

\section{Discussion\label{sec:discussion}}
The present study generalizes the results on the macroscopic friction of frictionless bead packs, previously obtained in simple shear, 
to other loading paths, and proposes a form of the failure criterion valid for arbitrary stress directions. This failure condition is somewhat different
from the  Mohr-Coulomb condition and best expressed in the Lade-Duncan form. As previously observed~\cite{PR08a}, despite rather strong finite size effects, 
the system is able to sustain finite stress deviators in the macroscopic geometric limit, in which the Lade-Duncan parameter, evaluated at 
$k_{\infty} = 27.22 \pm 0.02$, is to be regarded as a basic geometric property of disordered sphere assemblies. Changes of volume fraction $\Phi$ as
deviatoric stresses evolve from zero to yield threshold values are quite small and erratic
(in spite of a very slight tendency toward contractance under small deviator,
and to volume increase close to failure) and might be neglected, given statistical uncertainties, in a first approach. Thus
$\Phi$ remains approximately equal to the RCP value. All classical characterizations of packing geometry and force networks
by scalar or orientation-averaged variables, including
the distribution of normal forces, do not distinguish anisotropic equilibrium states from the initial isotropic structures equilibrated under hydrostatic pressure.

Thus the equilibrated configurations may be regarded as \emph{anisotropic random close-packing states}. 
Isotropic RCP states, in the limit of rigid particles, are local minima of sample volume in configuration space, under
the constraint of impenetrability of particles.
Anisotropic ones also minimize the potential energy of the applied stresses, \emph{viz.}
$$
W = -V\sum _{\alpha,\beta} \Sigma_{\alpha\beta} \epsilon_{\alpha\beta}
$$
where strain tensor $\ww{\epsilon}$, assumed small, has to be defined with respect to some arbitrary reference configuration. Consequently, they do not 
maximize volume fraction $\Phi$, and, although stable equilibrium states, do not qualify as ``strictly jammed'' according to the definition of Refs.~\cite{TOST01,DTSC04}.
That their volume fraction is no smaller (and occasionally slightly larger) than $\Phi_{\text{RCP}}$ obtained in isotropic configurations is due to the multiplicity
of different possible equilibrium networks and minima of potential energies $W$, which are not connected by quasistatic trajectories.

Fabric and force anisotropies can be 
efficiently characterized with one coefficient in an expansion in spherical harmonics. Each one of such coefficients is a function of stress anisotropy. 
The existence of such relations is specific to frictionless systems, 
in which any change of stress direction tends to entail rearrangements and changes in the contact network.
Meanwhile, like in granular systems with friction, stresses can be expressed, in good approximation, as combinations of fabric and force anisotropy parameters.  

Elastic moduli exhibit similar
anomalies in the rigid limit as in isotropic states, with a nearly uniaxial tensor of elastic moduli, the dominant eigenvalue of which (the only non-singular one) 
expresses the response to load increments parallel to the preexisting load in stress space. Meanwhile, the moduli in orthogonal directions vanish in the rigid
limit, as in isotropic systems (and the ``density of states'' for eigenmodes is expected to exhibit the same singularities~\cite{WNW05}). 
These properties can be expected to apply to any situation of very small force indeterminacy in particle packings. 

Our results seem to indicate that a deterministic stress-strain curve for monotonic loading along given deviatoric paths should be obtained 
in the macroscopic limit, thereby contradicting the conclusions of~\cite{CR2000}, based on an exactly rigid system in 2D, although the simulated samples
still seem too small to reach a clear conclusion about the regression of strain fluctuations for given applied stresses. This point obviously deserves further
investigations, as well as the spatial structure and displacement correlations in deformation and rearrangement mechanisms. The possibility of a diverging length
scale in the rigid limit of $\kappa\to +\infty$ 
(entailing the non-commutation of the limits of $\kappa\to +\infty$ and of $N\to +\infty$) should be explored in further
simulations of larger systems with varying stiffness level.

Another issue worth investigating is that of the possible uniqueness of equilibrium states, in the statistical sense, under a given supported state of stress. 
Just like simulation results appear to support the idea of a unique RCP state under isotropic pressure~\cite{iviso1}, provided a fast enough assembling process
bypasses crystal nucleation, the results reported here suggest that the internal state of the packing in equilibrium could be uniquely determined by the
current value of stresses, whatever the loading history. Such a conjecture is, in particular, supported by the observation of a one-to-one correspondence between 
stress and all measured internal state variables, such as fabric or force anisotropy parameters. 

Eventually, we expect that the knowledge of the behavior of frictionless granular assemblies will be useful in the design of compaction strategies (lubrication,
vibration, cyclic loading...), which can be regarded as methods to circumvent the influence of friction~\cite{iviso1}. Other interesting
perspectives involve the treatment of different particle shapes~\cite{Donev-ellipse,ARPS07} and polydispersities~\cite{VRDY07}.

\end{document}